\documentclass[a4paper,11pt]{article}
\usepackage{pos}
\usepackage{subcaption}

\title{Investigating double bump air showers with the SKA-Low}

\author*[d]{V.~De Henau}
\author[b]{S.~Bouma}
\author[c]{J.~Bray}
\author[d,e]{S.~Buitink}
\author[d,e]{A.~Corstanje}
\author[d]{M.~Desmet}
\author[f]{E.~Dickinson}
\author[e]{L.~van Dongen}
\author[g,h]{B.~Hare}
\author[i]{H.~He}
\author[e]{J.R.~H\"orandel}
\author[j,d]{T.~Huege}
\author[f]{C.W.~James}
\author[k,l]{M.~Jetti}
\author[b]{P.~Laub}
\author[j]{H.-J.~Mathes}
\author[e]{K.~Mulrey}
\author[b,m]{A.~Nelles}
\author[g]{O.~Scholten}
\author[h]{C.~Sterpka}
\author[h]{S.~ter Veen}
\author[b]{K.~Terveer}
\author[g,h]{P.~Turekova}
\author[o]{T.N.G.~Trinh}
\author[j,p]{S.~Saha}
\author[b]{S.~Sharma}
\author[c]{R.~Spencer}
\author[j]{D.~Veberi\v{c}}
\author[j]{K.~Watanabe}
\author[q]{M.~Waterson}
\author[r,s]{C.~Zhang}
\author[t]{P.~Zhang}
\author[i,u]{Y.~Zhang}

\abstract{\emph{Double-bump showers} are a rare class of extensive air showers (EAS) predicted by Monte Carlo simulations. They occur when a high-energy secondary particle, the leading particle, travels significantly farther than the rest, creating a distinct double-peaked longitudinal profile. So far, no experiment has been able to directly detect these showers. The unique radio footprint of double-bump showers, characterized by multiple Cherenkov rings, provides a way to reconstruct longitudinal profiles from radio observations. With its dense antenna array and broad frequency range, the Square Kilometer Array Observatory (SKAO) will be the first experiment capable of detecting these features, offering a new opportunity to probe hadronic interactions and constrain particle cross sections at high energies.

In our analysis, we simulate the EAS using CORSIKA with the CoREAS plugin for radio. We developed a new method based on the Akaike information criterion to identify double bump showers in simulations by analyzing their longitudinal profiles. Then we investigate the prevalence of these double bump showers across different cosmic ray primary particles and various hadronic interaction models. We create a skeleton of the EAS which consists of all the particles with at least $1\%$ of the primary energy, allowing us to confirm the leading particle hypothesis and track shower development following these particles. This will enable us to relate the attributes of the leading particle to measurable parameters. Depending on the exact shower properties, the radio footprint of a double bump shower can create a complex interference pattern, consisting of multiple rings. From this information, the longitudinal profiles can be extracted. SKA due to its dense antenna array and frequency range will be the first experiment able to observe these double bump showers in detail.}

\ConferenceLogo{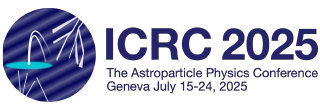}

\FullConference{39th International Cosmic Ray Conference (ICRC2025)\\
 15–24 July 2025\\
Geneva, Switzerland\\}

\begin{document}
\maketitle

\section{Introduction}
Most extensive air showers (EAS) exhibit a characteristic ``universal'' longitudinal development profile with a single, well-defined maximum. However, an intriguing class of events, known as double bump showers, deviates by exhibiting two distinct peaks in their longitudinal development. These anomalous profiles can provide valuable insight into the nature of the primary cosmic ray and its interaction processes in the atmosphere~\cite{Baus:2011kc}. Refining our understanding of such interactions at cosmic ray energies will enhance the accuracy of hadronic interaction models. In turn, this will improve the interpretation of extensive air shower (EAS) data, leading to better constraints on the composition and origin of cosmic rays, while also feeding back into the refinement of the interaction models themselves. Improved proton/helium separation is particularly promising for identifying the onset of the extragalactic component in the cosmic-ray spectrum. Furthermore, it may enable the detection of a helium-rich secondary galactic component, potentially originating from supernova remnants of Wolf-Rayet stars~\cite{Thoudam_2016}.

\begin{figure}
    \centering
    \includegraphics[width=\linewidth]{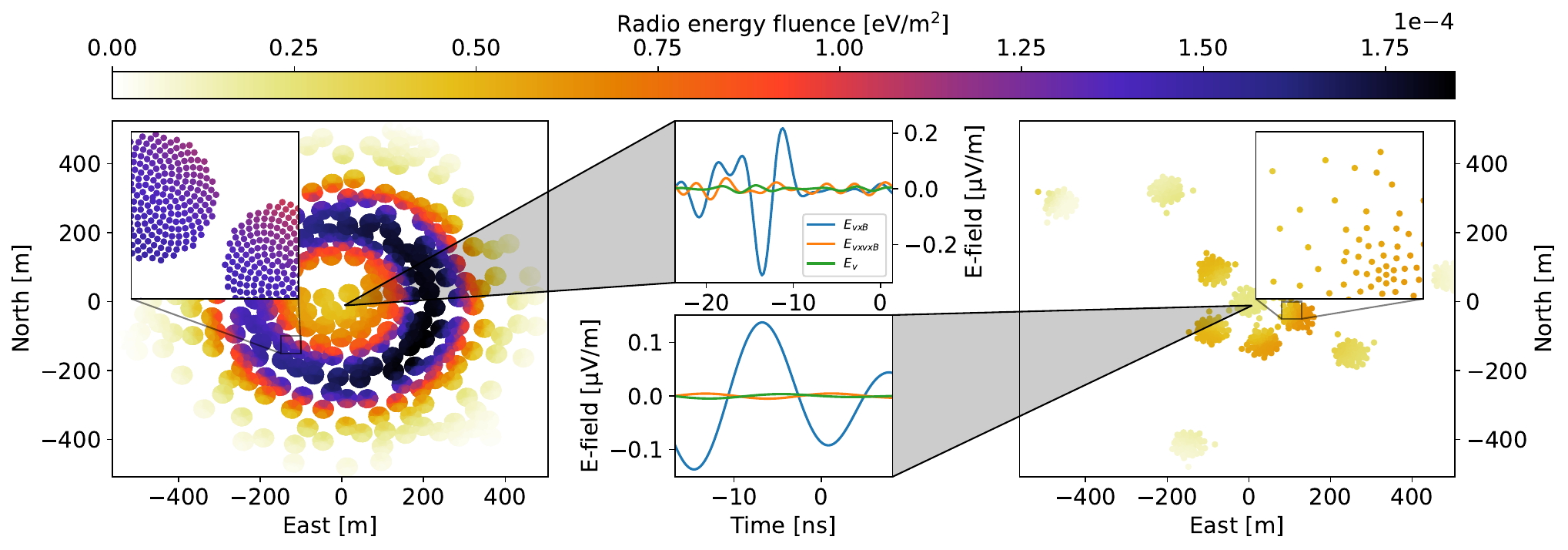}
    \caption{The same double bump EAS as it would be detected by SKA (left) or LOFAR (right) because individual showers will be observed with thousands of antennas distributed much more homogeneously. The complete radiation pattern immediately becomes apparent compared to LOFAR. Middle: example traces (in the shower-plane) of antennas at roughly the same position measured as by both SKA and LOFAR taking into account the different bandwidths, the double bump nature is clear from the unusual waveform shape in the trace signal of the SKA antenna.}
    \label{fig:LOFAR_and_SKA}
\end{figure}

Double bump showers are believed to occur when a secondary particle in the cascade travels significantly farther than others before initiating further interactions. The occurrence of such events is sensitive to the hadronic interaction model used in the simulation. Remarkably, the double-peak structure also manifests in the radio emission detected at ground level. However, only with the dense and homogeneous coverage of the SKA (bandwidth of $50$ to $350$\,MHz) as opposed to the more sparse LOFAR (bandwidth of $30$ to $80$\,MHz) configuration can this structure be clearly resolved. Fig.~\ref{fig:LOFAR_and_SKA} illustrates the radio footprint of a double bump event as detected by LOFAR (right) and SKA (left). In certain antennas only when looking with the SKA-Low bandwidth, the distinctive double-peak structure is directly observable in the time-domain traces as two separate radio pulses, revealing the anomalous nature of the shower.

The showers were simulated using CORSIKA~\cite{1998cmcc.book.....H} with the CoREAS~\cite{Huege_2013} for the radio emission using QGSJET (version: QGSJETII-04)~\cite{Ostapchenko_2006}, SIBYLL (SIBYLL 2.3d.)~\cite{PhysRevD.50.5710}, or EPOS  (version: EPOS-LHC)~\cite{WERNER200881} to simulate  the hadronic interactions plugin, using a star-shaped antenna pattern. A Fourier-based interpolation~\cite{Corstanje_2023} as implemented in the NuRadioReco software~\cite{Glaser_2019} was used to construct the radio footprint.

\section{Identifying double bumps}
In order to get the fraction of EAS's that are double bumps, we analyze the longitudinal profiles without simulating the radio emission to minimize computational time. The single-bump profile is modeled using the reduced Gaisser-Hillas function Eq.~\eqref{eq:GH_reduced}~\cite{1977ICRC....8..353G,ANDRINGA2011360}:
\begin{equation}
    N(X) = N_{\text{max}}\left(1 + R \frac{X - X_{\text{max}}}{L}\right)^{R^{-2}} e^{-\frac{X - X_{\text{max}}}{R ~ L}} \; .
    \label{eq:GH_reduced}
\end{equation}
The double-bump profile is modelled as the sum of two such functions Eq.~\eqref{eq:GH_reduced}. 
While a previous study identified double bump events based on a set of predefined rules depending on the parameters from the resulting double bump fit~\cite{Baus:2011kc}, we instead apply the Akaike Information Criterion (AIC)~\cite{1311138} to evaluate the relative quality of the two models. The AIC balances the goodness-of-fit with the model complexity and is given by $\mathrm{AIC} = 2k - 2 \ln(L) = 2k+\chi^2$. Where $k$ is the number of model parameters and  $L$ is the maximum likelihood. A lower AIC value indicates a better model.

\begin{figure}
     \centering
     \begin{subfigure}[b]{0.3\textwidth}
         \centering
         \includegraphics[width=\textwidth]{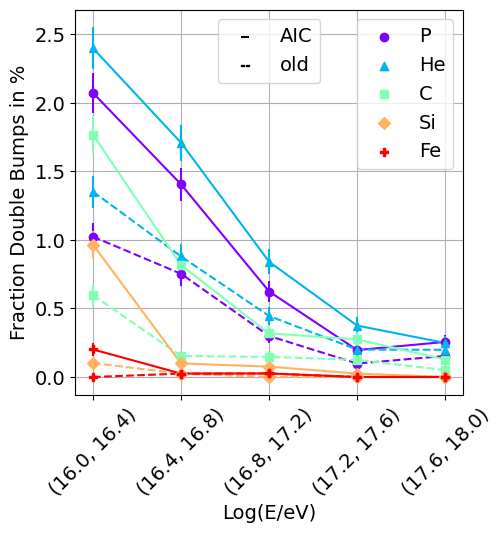}
         \caption{QGSJETII-04}
         \label{fig:plot_Double_Bump_QGSII_Energy_Double_Bump}
     \end{subfigure}
     \hfill
     \begin{subfigure}[b]{0.3\textwidth}
         \centering
         \includegraphics[width=\textwidth]{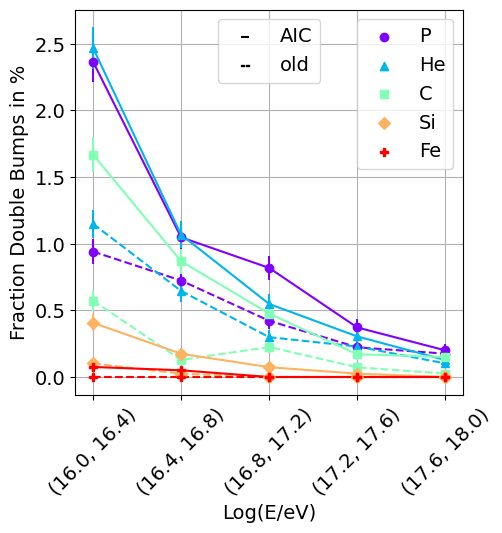}
         \caption{EPOS-LHC}
         \label{fig:plot_Double_Bump_EPOS_Energy_Double_Bump}
     \end{subfigure}
     \hfill
     \begin{subfigure}[b]{0.3\textwidth}
         \centering
         \includegraphics[width=\textwidth]{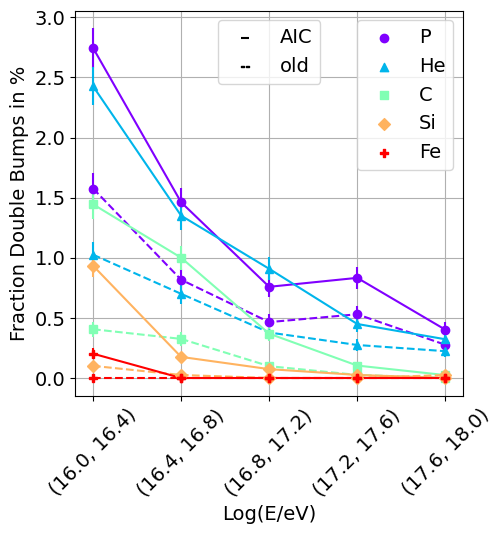}
         \caption{SIBYLL 2.3d.}
         \label{fig:plot_Double_Bump_SIBYLL_Energy_Double_Bump}
     \end{subfigure}
        \caption{Fraction of anomalous shower profiles for different primaries for different hadronic interaction models, plotted against $\lg(E/\text{eV})$. Each data point represents the percentage of anomalous shower profiles with double bumps detected for different primary particles (proton, helium, carbon, silicon, iron).}
        \label{fig:plot_double_bumps_Energy}
\end{figure}

Applying these methods, we estimate the fraction of anomalous showers in simulated samples for primary energies from $10^{16}$ to $10^{18}$\,eV (zenith angle of $15$\textdegree to $45$ \textdegree). Each primary species (proton, helium, carbon, silicon, iron) was simulated with $20\,000$ events. Fig.~\ref{fig:plot_double_bumps_Energy} shows the frequency of double bump events for different primaries and hadronic interaction models, for the existing method (old) as described in Ref.~\cite{Baus:2011kc} and using the AIC method. Notably, helium and proton primaries exhibit comparable rates of anomalous profiles. The overall occurrence of double bump profiles decreases with increasing energy, consistent with expectations based on rising interaction cross-sections.

\section{Looking inside EAS}
To investigate the internal structure of EAS, we modified the CORSIKA simulation to access the energy and trajectory of individual secondary particles. The example showers used in the rest of these proceedings were simulated with QGSJETII-04 and $\theta =60$\textdegree at $10^{15}$\,eV so that we have a higher amount of double bumps and the shower development usually completes before hitting the ground.

\begin{figure}
    \centering
    \includegraphics[width=\linewidth]{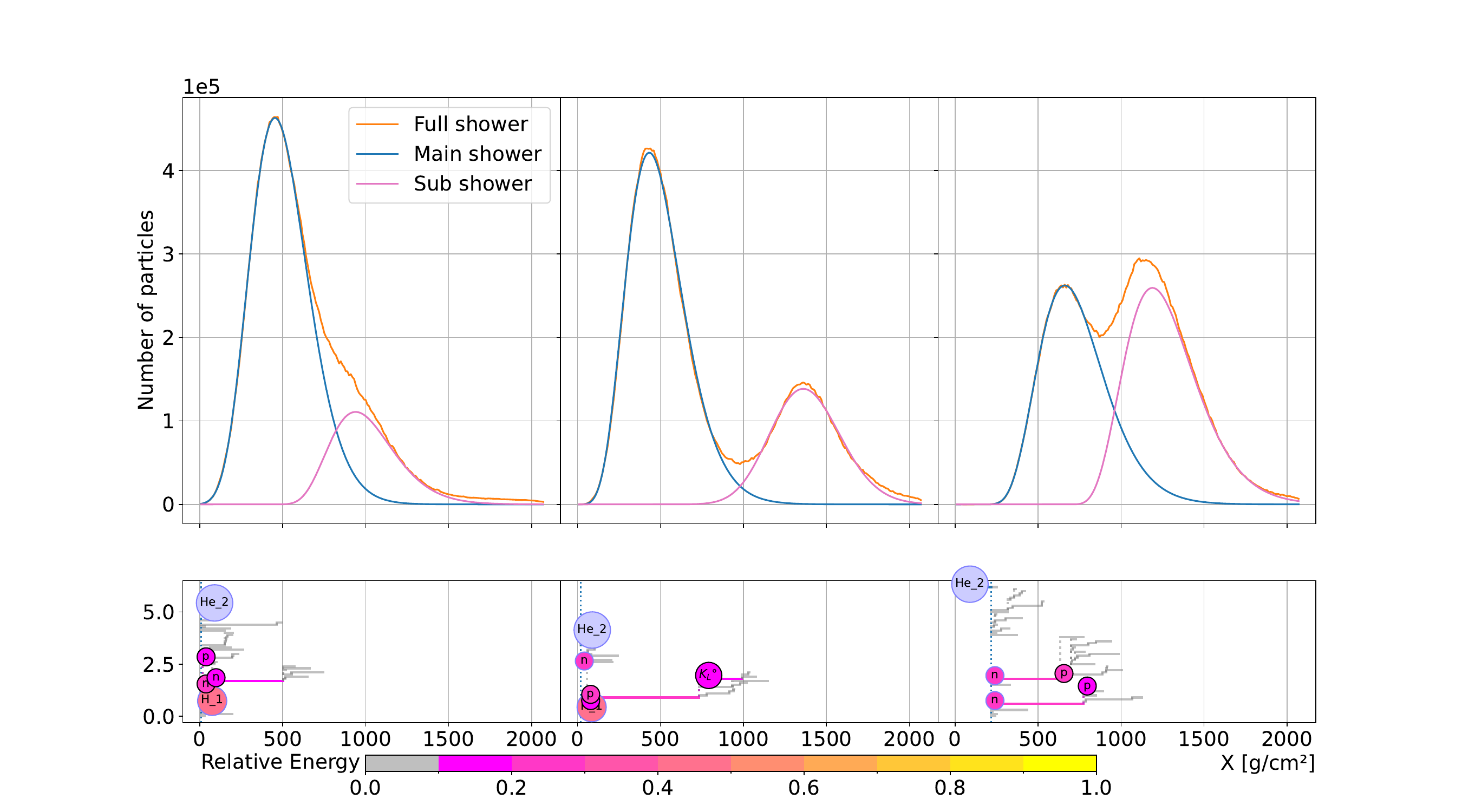}
    \caption{Longitudinal profile of a double bump shower with the path travelled by all particles which have more than $1\%$ of the primaries energy. Left: The sub bump is caused by a negatively neutron penetrating deeply in the atmosphere, Single leading particle case. Middle: The sub bump is caused by positive pion followed by a negative kaon, Ladder case. Right: The sub bump is caused by two neutrons travelling parallel, Branching case.}
    \label{fig:Skeleton_shower_DBs}
\end{figure}

We applied an energy cut of $E > 0.1 \times E_{\text{primary}}$ to isolate high-energy secondaries responsible for significant shower development. This allowed us to construct skeleton plots, as shown in Fig.~\ref{fig:Skeleton_shower_DBs}. Skeleton plots illustrate the correlation between the longitudinal development of an EAS and the propagation of energetic secondaries. The top panel in Fig.~\ref{fig:Skeleton_shower_DBs} shows the standard longitudinal profile, while the lower panel visualizes the tracked particle paths and their significant interactions. These visualizations are crucial for identifying candidates for the so-called leading particle -- a particle hypothesized to traverse large distances before initiating a secondary cascade that causes the second bump.

The probability for a particle penetrating a slant depth greater than $\Delta X$ is given by: $P(\Delta X) = e^{-\Delta X / \lambda}$, with $\lambda$ denoting the hadronic interaction length in air. As $\Delta X$ cannot be measured directly, we use the distance between the two shower maxima, $\Delta X_{\text{max}}$, as a proxy.
To validate this we need to identify the leading particle, we search for the secondary $i$ that minimizes both $\Delta X_i \sim \Delta X_{\text{max}}$ and $N_{\text{max}} / N_{\text{max}} \sim E_{i} / (E_{\text{primary}}-E_{i})$.
Our analysis of multiple skeleton plots reveals that the single-particle explanation is not always sufficient. We identify three formation mechanisms for double bump showers:
\begin{itemize}
    \item Single leading particle double bump: A single secondary particle travels deep before initiating the second cascade.
    \item Ladder double bump: A cascade of two secondaries particles, each penetrating successively deeper.
    \item Branching double bump: Two independent, energetic secondaries each contribute significantly to the second bump.
\end{itemize}
These examples of these scenarios are depicted in Fig.~\ref{fig:Skeleton_shower_DBs}.

To differentiate between these types, we developed a classification algorithm. This algorithm systematically evaluates the trajectories and energy deposits of high-energy particles in the shower to identify the most plausible underlying scenario that could have generated the observed longitudinal double bump profile.

\begin{figure}
    \centering
    \includegraphics[width=\linewidth]{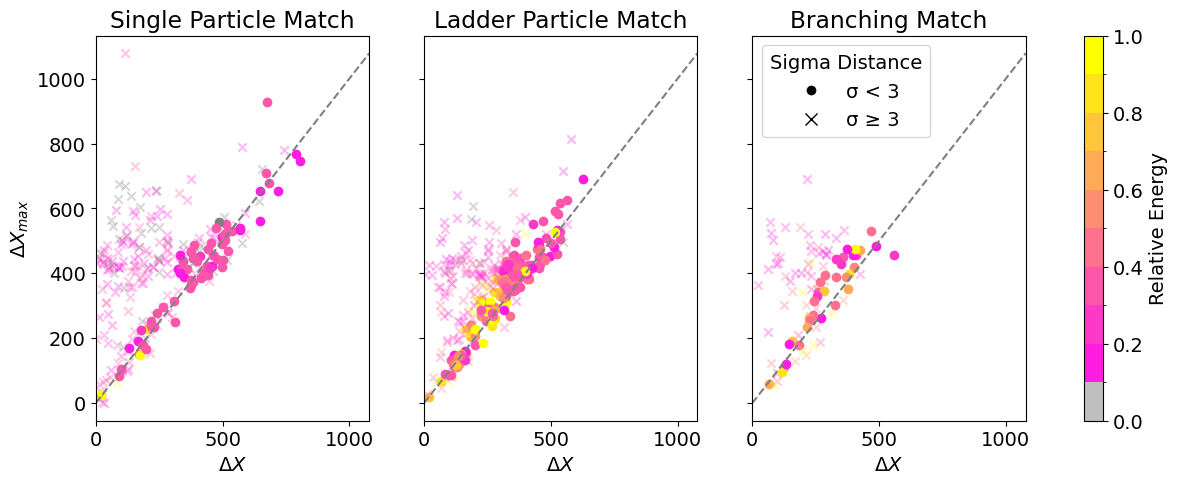}
    \caption{The different type of causes for double bumps: Single leading particle (single particle penetrating deeply), Ladder (successive penetration of particles together traveling a significant distance) and Branching (independent deeply penetrating particles traveling significant distances).
    A clear linear correlation is observed between $\Delta X$ and $\Delta X_{\text{max}}$ for events within $3\sigma$, as expected. Outliers from this trend are likely cases where the AIC method misidentifies showers as double-bump events, despite the underlying particle content suggesting otherwise.}
    \label{fig:Double_bump_types}
\end{figure}

The distinction between these two scenarios is illustrated in Fig.~\ref{fig:Double_bump_types}, which shows the correlation between the algorithm-inferred propagation distances ($\Delta X$) and the observed separation between the two peaks in the longitudinal profile ($\Delta X_{\text{max}}$). We differentiate between cases where the algorithm found a match within $3\sigma$ (circles) and those where it did not (crosses). The latter likely represent false positives instances where the AIC-based double bump identification algorithm indicated a double structure, but no corresponding signature is evident in the particle level skeleton. In contrast, the strong correlation observed in the $3\sigma$ (the uncertainties stem forth from the fitting of the main and sub bump in the longitudinal profile) matches supports the interpretation that specific particle-level processes underlie the observed double bump structure.

\section{Reconstructing the slant depth of both bumps}

\begin{figure}
    \centering
    \includegraphics[width=\linewidth]{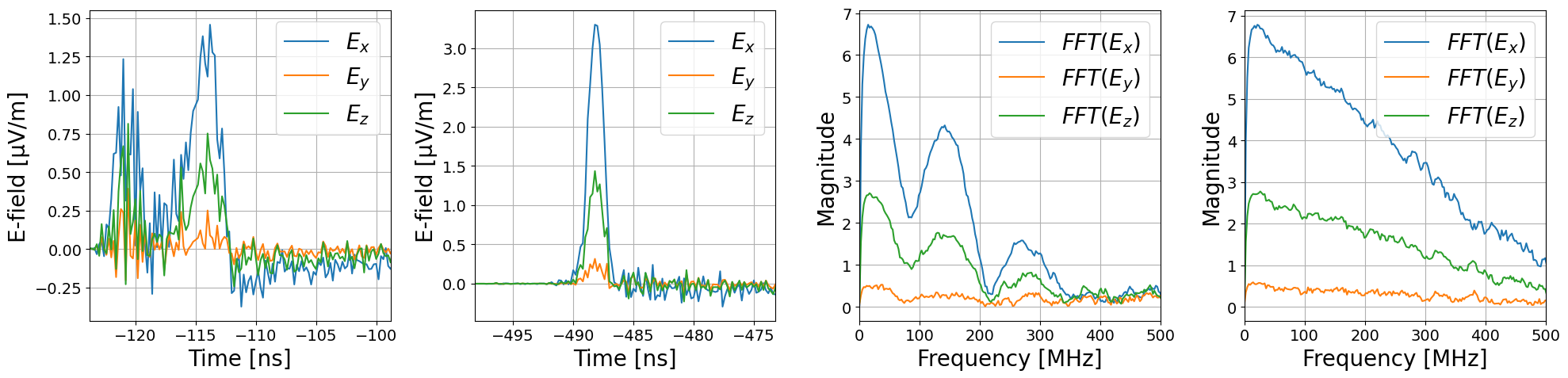}
    \caption{Left: Time-domain electric field traces for two antennas at different lateral distances from the shower core (at 50\,m and at 200\,m). Right: Corresponding Fourier transforms of the electric field components. A frequency dip around $100$\,MHz is clearly visible for the closer antenna.}
    \label{fig:Traces_and_fft}
\end{figure}

\begin{figure}
     \centering
     \begin{subfigure}[b]{0.49\textwidth}
         \centering
         \includegraphics[width=\textwidth]{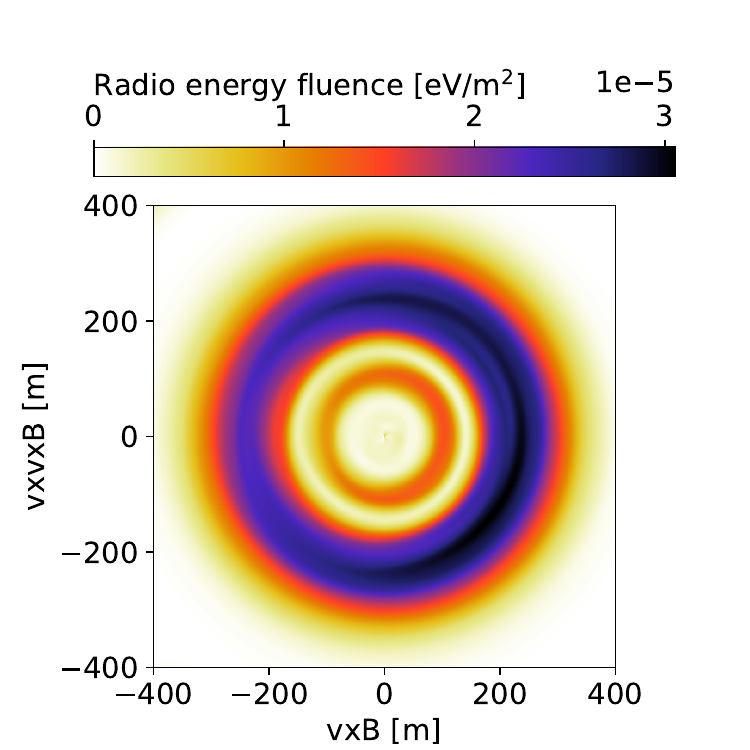}
     \end{subfigure}
     \hfill
     \begin{subfigure}[b]{0.49\textwidth}
         \centering
         \includegraphics[width=\textwidth]{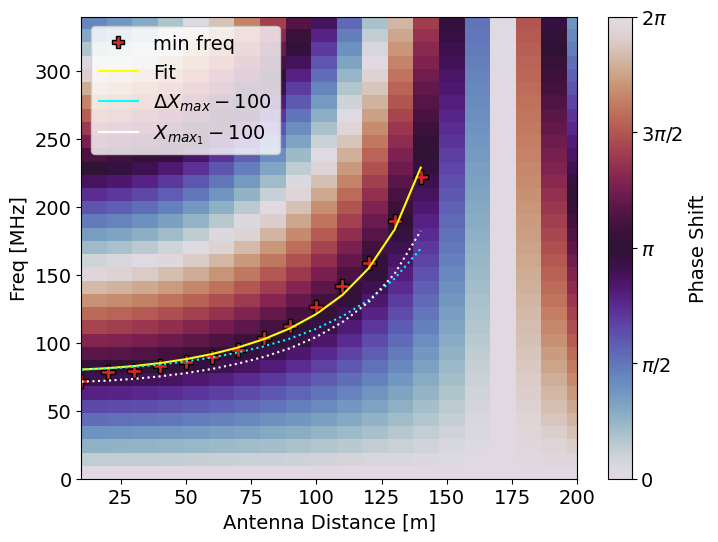} 
     \end{subfigure}
     \caption{Left: Example radio fluence plot in the shower plane of a double bump shower filtered between $200$ and $250$\,MHz. Showing a second ring inside the main Cherenkov ring, caused by the second bump.
     Right: Reconstructed dip frequencies along the positive $v \times B$ axis. Red crosses indicate the extracted dip positions for each antenna, while the yellow curve shows the best-fit model. The background heatmap displays the predicted phase difference $\Delta \phi = \Delta T(\vec{r}) \times \omega$ as a function of position and frequency. Blue dotted lines illustrate the effect of increasing the separation $\Delta X_{\text{max}}$ by $100$\,g/cm$^2$, while the white dotted line shows the impact of shifting only $X_{\text{max}_1}$.}
     \label{fig:Fluence_and_dip}
\end{figure}

Examining the radio traces of a double bump shower (Fig.~\ref{fig:Skeleton_shower_DBs}, middle) for antennas along the $v \times B$ axis reveals a clear signature of the structure (Fig.~\ref{fig:Traces_and_fft}, left). For the antenna located $50$ m from the shower core, the double bump nature is evident, with two distinct peaks, the earlier one corresponding to the second bump. In contrast, at a distance of $200$\,m, this double peak structure is no longer visible in the time trace, likely due to geometric or coherence effects.
When analyzing the corresponding Fourier transforms, we observe that traces from antennas closer to the core exhibit a distinct dip in the frequency amplitude spectrum (Fig.~\ref{fig:Traces_and_fft}, right). This feature is only present for antennas located within the outermost Cherenkov ring, suggesting that it is a direct consequence of the double bump structure in the shower development. Further supporting this interpretation, the interpolated radio fluence footprint filtered between $200$ and $250$\,MHz (Fig.~\ref{fig:Fluence_and_dip}, left) shows a secondary, smaller ring, which we attribute to the second bump in the longitudinal profile.

We hypothesize that the dip in the frequency spectrum is caused by destructive interference between the signals originating from the two shower maxima. To test this, we model each $X_{\text{max}}$ as a point-like emitter and compute the time delay $\Delta T$ between their respective signals as received at a given antenna location. The condition for destructive interference corresponds to a phase difference of $\pi$ between the two signals. Given a time delay $\Delta T$, the associated frequency where this condition is met is derived in Eq.~\eqref{eq:f}.
\begin{align}
  \Delta T(\Vec{r}) = \frac{\Delta \phi}{\omega}
    \begin{cases}
      \Delta \phi = 0~~~\text{const.}~(\text{mod}~2\pi) \\
      \Delta \phi = \pi~~~\text{destr.}~(\text{mod}~2\pi) \\
    \end{cases}
    \label{eq:f}
\end{align}
Using this we can predict the frequencies at which destructive interference occurs for each antenna. These predicted dip frequencies align well with the observed dips, if they occur above roughly $200$\,MHz. Beyond these ranges, discrepancies arise, likely due to additional effects not accounted for in the current model, for the same reason the second dip does not occur at the expected location predicted by choosing  $\Delta \phi = +2\pi$.
This allows us to invert the method: by identifying the dip frequencies in the FFT traces, we can fit for the positions of the two shower maxima, $X_{\text{max}_1}$ and $X_{\text{max}_2}$. 
Looking at Fig. ~\ref{fig:Fluence_and_dip}, right,  shows the reconstructed dip frequencies for antennas along the positive $v \times B$ axis. Red crosses mark the dip, and the yellow line is a best-fit curve. The background heatmap represents the predicted phase difference for a given antenna position and frequency. The blue lines illustrate the sensitivity of nearby antennas to the separation $\Delta X_{\text{max}}$: shifting $\Delta X_{\text{max}}$ by $100$\,g/cm$^2$ results in a noticeable change that aligns with the fit. In contrast, the white lines correspond to a shift in $X_{\text{max}_1}$ alone, which causes a different pattern, confirming that nearby antennas are more sensitive to the separation than the absolute depth.

Using this proof-of-principle reconstruction method, we obtain 
$X_{\text{max}_1} = 685.09$\,g/cm$^2$ and $X_{\text{max}_2} = 1484.62$\,g/cm$^2$. For comparison, the AIC-based method yields 
$X_{\text{max}_1} = 659.93$\,g/cm$^2$ and $X_{\text{max}_2} = 1496.37$\,g/cm$^2$, indicating that at least one of the two shower maxima was reconstructed within $30$\,g/cm$^2$ in both approaches.

To further interpret these results, it is necessary to estimate the energy of both the primary and leading secondary particles, quantities which are related to $N_{\text{max}_1}$ and $N_{\text{max}_2}$, respectively. Two possible approaches could be pursued: (1) decompose the radio footprint into the contributions from each of the two shower components, or (2) use the reconstructed values of $X_{\text{max}_1}$ and $X_{\text{max}_2}$ in conjunction with fast radio simulation tools such as \texttt{SMIET} \cite{desmet2025smietfastaccuratesynthesis} or MGMR~\cite{Mitra_2023}, to determine the corresponding $N_{\text{max}_1}$ and $N_{\text{max}_2}$ values that best match the measured signal. This demonstrates that the signal measured by SKA-Low can be used to reconstruct double-peaked air shower profiles.

\bibliographystyle{ICRC}
\bibliography{refs}

\newpage
\section*{Affiliations}
\scriptsize
\noindent
$^{a}$ Universit\'e Libre de Bruxelles, Science Faculty CP230, B-1050 Brussels, Belgium\\
$^{b}$ ECAP, Friedrich-Alexander-Universität Erlangen-Nürnberg, 91058 Erlangen, Germany \\
$^{c}$ Jodrell Bank Centre for Astrophysics, Dept.\ of Physics \& Astronomy, University of Manchester, UK \\
$^{d}$ Vrije Universiteit Brussel, Inter-University Institute For High Energies (IIHE), Pleinlaan~2, 1050 Brussels, Belgium \\
$^{e}$ Department of Astrophysics/IMAPP, Radboud University Nijmegen, The Netherlands \\
$^{f}$ International Centre for Radio Astronomy Research, Curtin University, Bentley, 6102, WA, Australia \\
$^{g}$ Kapteyn Astronomical Institute, University of Groningen, Netherlands \\
$^{h}$ Netherlands Institute for Radio Astronomy (ASTRON), Dwingeloo, The Netherlands \\
$^{i}$ Key Laboratory of Dark Matter and Space Astronomy, Purple Mountain Observatory, Chinese Academy of Sciences, No.~10 Yuanhua Road, Nanjing, China \\
$^{j}$ Institut f\"ur Astroteilchenphysik, Karlsruhe Institute of Technology (KIT), Germany \\
$^{k}$ Max-Planck Institut für Astrophysik, Karl-Schwarzschild-Str. 1, 85748 Garching, Germany \\
$^{l}$ Ludwig-Maximilians-Universität München (LMU), München, Germany \\
$^{m}$ Deutsches Elektronen-Synchrotron DESY, Platanenallee 6, 15738 Zeuthen, Germany \\
$^{n}$ Department of Physics, Khalifa University, P.O. Box 127788, Abu Dhabi, United Arab Emirates \\
$^{o}$ Physics Education Department, School of Education, Can Tho University, Campus II, 3/2 Street, Ninh Kieu District, Can Tho City, Viet Nam \\
$^{p}$ Department of Physics, Indian Institute of Technology Kanpur, Kanpur, UP-208016, India \\
$^{q}$ SKA Observatory, Jodrell Bank, Lower Withington, Macclesfield, SK11 9FT, UK \\
$^{r}$ School of Astronomy and Space Science, Nanjing University, Nanjing 210023, China \\
$^{s}$ Key laboratory of Modern Astronomy and Astrophysics, Nanjing University, Ministry of Education, Nanjing 210023, China \\
$^{t}$ School of Electronic Engineering, Xidian University, No.2 South Taibai Road, Xi’an, China \\
$^{u}$ School of Astronomy and Space Science, University of Science and Technology of China, Hefei 230026, China \\

\section*{Acknowledgments}
SBo, AN and KT acknowledge funding through the Verbundforschung of the German Federal Ministry of Research, Technology and Space (BMFTR). PL, KW and MJ are supported by the Deutsche Forschungsgemeinschaft (DFG, German Research Foundation) – Projektnummer 531213488. MD is supported by the Flemish Foundation for Scientific Research (FWO-AL991). ST acknowledges funding from the Khalifa University RIG-S-2023-070 grant. SB acknowledges funding from the Medium-Scale Infrastructure program of the Flemish Foundation for Scientific Research (FWO). KM acknowledges funding from the Netherlands Research School for Astronomy (NOVA) Phase 6 Instrumentation Call. The authors gratefully acknowledge the computing time provided on the high-performance computer HoreKa by the National High-Performance Computing Center at KIT (NHR@KIT). This center is jointly supported by the Federal Ministry of Education and Research and the Ministry of Science, Research and the Arts of Baden-Württemberg, as part of the National High-Performance Computing (NHR) joint funding program. HoreKa is partly funded by the German Research Foundation.

\end{document}